\documentclass[12pt,a4paper]{article}
\usepackage{amsmath}
\usepackage{latexsym}
\makeatletter
\def\rddots{\mathinner{\mkern1mu\raise\p@%
    \vbox{\kern7\p@\hbox{.}}\mkern2mu%
    \raise4\p@\hbox{.}\mkern2mu\raise7\p@\hbox{.}\mkern1mu}}
\makeatother
\begin{document}

\title{\sl A Generalization of the Lamb--Bateman 
Integral Equation and Fractional Derivatives : A Comment}
\author{
  Kazuyuki FUJII
  \thanks{E-mail address : fujii@yokohama-cu.ac.jp }\\
  ${}^{*}$Department of Mathematical Sciences\\
  Yokohama City University\\
  Yokohama, 236--0027\\
  Japan
  }
\date{}
\maketitle
\begin{abstract}
  In this note a generalization of the Lamb--Bateman integral equation 
is presented and its solution is given in terms of {\bf fractional derivatives}. 
This is a comment one to the paper by Babusci, Dattoli and 
Sacchetti (arXiv:1006.0184 [math-ph]).
\end{abstract}

In the paper \cite{BDS} the authors reconsidered the integral equation 
presented by Lamb \cite{HL}
\begin{equation}
\label{eq:lamb}
\int_{0}^{\infty}u(x-y^{2})dy=f(x),
\end{equation}
where $f(x)$ is a function given and $u(x)$ is a function to be 
determined. First, an interesting solution
\begin{equation}
\label{eq:bateman}
u(x)=\frac{2}{\pi}\int_{-\infty}^{x}\frac{f^{\prime}(\xi)}{\sqrt{x-\xi}}d\xi
\end{equation}
was given by Harry Bateman. Note that this form is a bit different from 
the original one. However, the proof by Bateman himself has not been known. 
In \cite{BDS} a modern derivation in terms of {\bf fractional derivatives} was 
given to be
\begin{equation}
\label{eq:solution}
u(x)=\frac{2}{\sqrt{\pi}}{\partial_{x}}^{1/2}f(x)=
\frac{2}{\pi}\int_{-\infty}^{x}\frac{f^{\prime}(\xi)}{\sqrt{x-\xi}}d\xi,
\end{equation}
which is instructive enough. Here we set $\partial_{x}=\frac{d}{dx}$ for 
simplicity. In the following this notation is used.

We would like to generalize the equation (\ref{eq:lamb}). In \cite{BDS} 
a generalization was presented to be
\begin{equation}
\label{eq:bds}
\int_{0}^{\infty}u(x-y^{m})dy=f(x)\quad (m\geq 2).
\end{equation}
However, we present another generalization. By rewritting 
(\ref{eq:lamb}) to be 
\[
\int_{-\infty}^{\infty}u(x-y^{2})dy=f(x),
\]
(the coefficient $1/2$ has been omitted for simplicity) 
we give a multi--dimensional integral equation
\begin{equation}
\label{eq:fujii}
\int\int{\cdots}\int_{{\bf R}^{n}}u(x-\sum_{j=1}^{n}{y_{j}}^{2})
dy_{1}dy_{2}{\cdots}dy_{n}=f(x),
\end{equation}
where the function $f$ is assumed to be ``good" (which means 
that $f$ has all properties required in the process of calculation).

Let us solve the equation in a similar way as in \cite{BDS}. 
The polar coordinates of ${\bf R}^{n}$
\[ 
{\bf r}=(r,\theta_{n-2},\cdots,\theta_{2},\theta_{1},\phi)
\ \longrightarrow\ 
{\bf y}=(y_{1},y_{2},y_{3},\cdots,y_{n-1},y_{n})
\]
is given by 
($0\leq r <\infty,\ 0\leq \theta_{n-2},\cdots,\theta_{2},\theta_{1}\leq \pi,\ 
0\leq \phi <2\pi$)
\begin{eqnarray}
y_{n}&=&r\cos\theta_{n-2} \nonumber \\
y_{n-1}&=&r\sin\theta_{n-2}\cos\theta_{n-3} \nonumber \\
&\vdots& \nonumber \\
y_{3}&=&r\sin\theta_{n-2}\cos\theta_{n-3}\cdots\sin\theta_{2}\cos\theta_{1}
\nonumber \\
y_{2}&=&r\sin\theta_{n-2}\cos\theta_{n-3}\cdots\sin\theta_{2}\sin\theta_{1}\sin\phi
\nonumber \\
y_{1}&=&r\sin\theta_{n-2}\cos\theta_{n-3}\cdots\sin\theta_{2}\sin\theta_{1}\cos\phi.
\end{eqnarray}
The Jacobian of this coordinate transformation is given by
\begin{equation}
{\bf J}
=\det\left(\frac{\partial {\bf y}}{\partial {\bf r}}\right)
=\pm r^{n-1}\sin^{n-2}(\theta_{n-2})\sin^{n-3}(\theta_{n-3})\cdots \sin\theta_{1}.
\end{equation}
Note that the proof is not so easy for undergraduates.

Then, under the coordinate transformation the equation (\ref{eq:fujii}) becomes
\begin{equation}
\label{eq:fujii-2}
\mbox{Vol}(S^{n-1})\int_{0}^{\infty}r^{n-1}u(x-r^{2})dr=f(x)
\end{equation}
where $\mbox{Vol}(S^{n-1})$ is the volume of the (n-1)--dimensional sphere 
$S^{n-1}$ given by
\begin{eqnarray*}
\mbox{Vol}(S^{n-1})
&=&
\int_{0}^{\pi}\sin^{n-2}(\theta_{n-2})d\theta_{n-2}
\int_{0}^{\pi}\sin^{n-3}(\theta_{n-3})d\theta_{n-3}
\cdots
\int_{0}^{\pi}\sin\theta_{1}d\theta_{1}
\int_{0}^{2\pi}1d\phi  \\
&=&
B\left(\frac{n-1}{2},\frac{1}{2}\right)
B\left(\frac{n-2}{2},\frac{1}{2}\right)
\cdots
B\left(\frac{3}{2},\frac{1}{2}\right)
B\left(\frac{2}{2},\frac{1}{2}\right)\times 2\pi  \\
&=&
2\pi
\frac{\Gamma\left(\frac{n-1}{2}\right)\Gamma\left(\frac{1}{2}\right)}
{\Gamma\left(\frac{n}{2}\right)}
\frac{\Gamma\left(\frac{n-2}{2}\right)\Gamma\left(\frac{1}{2}\right)}
{\Gamma\left(\frac{n-1}{2}\right)}
\cdots
\frac{\Gamma\left(\frac{3}{2}\right)\Gamma\left(\frac{1}{2}\right)}
{\Gamma\left(\frac{4}{2}\right)}
\frac{\Gamma\left(\frac{2}{2}\right)\Gamma\left(\frac{1}{2}\right)}
{\Gamma\left(\frac{3}{2}\right)}  \\
&=&
2\pi
\frac{\Gamma\left(\frac{1}{2}\right)^{n-2}}{\Gamma\left(\frac{n}{2}\right)}  \\
&=&\frac{2\pi^{\frac{n}{2}}}{\Gamma\left(\frac{n}{2}\right)},
\end{eqnarray*}
where the calculation is based on the following properties
\[
\int_{0}^{\pi}\sin^{m}\theta d\theta=B(\frac{m+1}{2},\frac{1}{2})
\]
and
\[
B(p,q)=\frac{\Gamma(p)\Gamma(q)}{\Gamma(p+q)}; 
\quad 
\Gamma(1)=1,\ \Gamma(1/2)=\sqrt{\pi}
\]
where $B(p,q)$ and $\Gamma(p)$ are respectively 
the Beta--function and Gamma--function given by
\[
B(p,q)=\int_{0}^{1}x^{p-1}(1-x)^{q-1}dx,
\quad
\Gamma(p)=\int_{0}^{\infty}e^{-x}x^{p-1}dx.
\]

Tentatively, by setting $C=\mbox{Vol}(S^{n-1})$ we have
\begin{equation}
\label{eq:fujii-3}
\int_{0}^{\infty}r^{n-1}u(x-r^{2})dr=\frac{1}{C}f(x).
\end{equation}
In terms of the formula of Taylor expansion (which is formal)
\[
e^{a\partial_{x}}g(x)=g(x+a)
\]
the left hand side of (\ref{eq:fujii-3}) becomes
\[
\mbox{LHS}=\int_{0}^{\infty}r^{n-1}e^{-r^{2}\partial_{x}}u(x)dr
=
\left\{\int_{0}^{\infty}r^{n-1}e^{-r^{2}\partial_{x}}dr\right\}u(x),
\]
so we set $a=\partial_{x}$ {\bf formally} and calculate the 
integral
\[
\int_{0}^{\infty}r^{n-1}e^{-r^{2}a}dr=
\int_{0}^{\infty}r^{n-1}e^{-ar^{2}}dr.
\]
It is easily performed by setting  
$t=ar^{2}\ (\Rightarrow r=(t/a)^{1/2})$ and becomes
\begin{eqnarray*}
\int_{0}^{\infty}r^{n-1}e^{-ar^{2}}dr
&=&
\int_{0}^{\infty}\left(\frac{t}{a}\right)^{\frac{n-1}{2}}
e^{-t}\frac{dt}{2\sqrt{at}} \\
&=&
\frac{1}{2}\left(\frac{1}{a}\right)^{\frac{n}{2}}
\int_{0}^{\infty}t^{\frac{n}{2}-1}e^{-t}dt \\
&=&
\frac{1}{2}\Gamma\left(\frac{n}{2}\right)a^{-\frac{n}{2}}.
\end{eqnarray*}
Here, by inserting $\partial_{x}$ in place of $a$
\[
\left\{
\frac{1}{2}\Gamma\left(\frac{n}{2}\right){\partial_{x}}^{-\frac{n}{2}}
\right\}u(x)
=
\frac{1}{C}f(x)
\]
and we obtain the (formal) solution
\begin{equation}
\label{eq:fujii-4}
u(x)=
\frac{2}{C\Gamma\left(\frac{n}{2}\right)}{\partial_{x}}^{\frac{n}{2}}f(x).
\end{equation}
The explicit value of $C$ gives
\[
C=\frac{2\pi^{\frac{n}{2}}}{\Gamma\left(\frac{n}{2}\right)}
\ \Longrightarrow\ 
\frac{2}{C\Gamma\left(\frac{n}{2}\right)}=\pi^{-\frac{n}{2}},
\]
so we have a good--looking form
\begin{equation}
\label{eq:fujii-5}
u(x)=\pi^{-\frac{n}{2}}{\partial_{x}}^{\frac{n}{2}}f(x).
\end{equation}

Now we divide $n$ into two cases :

\vspace{3mm}\noindent
(a)\ ${n=2m}$
\begin{equation}
\label{eq:fujii-6}
u(x)=\pi^{-m}{\partial_{x}}^{m}f(x)=\pi^{-m}f^{(m)}(x).
\end{equation}
In this case there is no problem.

\vspace{3mm}\noindent
(b)\ ${n=2m+1}$\ \ In the case
\[
u(x)=\pi^{-(m+\frac{1}{2})}{\partial_{x}}^{m+\frac{1}{2}}f(x),
\]
so we encounter the fractional derivatives. Since
\begin{eqnarray*}
u(x)
&=&\pi^{-(m+\frac{1}{2})}{\partial_{x}}^{m+\frac{1}{2}}f(x) \\
&=&\pi^{-(m+\frac{1}{2})}{\partial_{x}}^{-\frac{1}{2}+m+1}f(x) \\
&=&\pi^{-(m+\frac{1}{2})}{\partial_{x}}^{-\frac{1}{2}}
{\partial _{x}}^{m+1}f(x) \\
&=&\pi^{-(m+\frac{1}{2})}{\partial_{x}}^{-\frac{1}{2}}f^{(m+1)}(x),
\end{eqnarray*}
we can use the following well--known formula \cite{OS}, \cite{AA}
\begin{equation}
\label{eq:formula}
{\partial_{x}}^{-\mu}g(x)=
\frac{1}{\Gamma(\mu)}\int_{-\infty}^{x}
g(\xi)(x-\xi)^{\mu-1}d\xi
\quad (\mu>0).
\end{equation}

\noindent
By inserting $\mu=\frac{1}{2}$ into the formula 
we obtain ($\Gamma(1/2)=\sqrt{\pi}$)
\begin{eqnarray}
\label{eq:fujii-7}
u(x)
&=&\pi^{-(m+\frac{1}{2})}\frac{1}{\Gamma(\frac{1}{2})}\int_{-\infty}^{x}
f^{(m+1)}(\xi)(x-\xi)^{\frac{1}{2}-1}d\xi
\nonumber \\
&=&\pi^{-(m+1)}\int_{-\infty}^{x}\frac{f^{(m+1)}(\xi)}{\sqrt{x-\xi}}d\xi.
\end{eqnarray}

\vspace{5mm}\noindent
\begin{center}
\begin{Large}
{\bf Summary}
\end{Large}
\end{center}

\vspace{3mm}\noindent
{\bf Equation} :
\[
\int\int{\cdots}\int_{{\bf R}^{n}}u(x-\sum_{j=1}^{n}{y_{j}}^{2})
dy_{1}dy_{2}{\cdots}dy_{n}=f(x).
\]

\vspace{3mm}\noindent
{\bf Solution} :

(a)\ ${n=2m}$
\[
\ \ u(x)=\pi^{-m}f^{(m)}(x).
\]

(b)\ ${n=2m+1}$
\[
u(x)=\pi^{-(m+1)}\int_{-\infty}^{x}\frac{f^{(m+1)}(\xi)}{\sqrt{x-\xi}}d\xi.
\]

\vspace{10mm}
A comment is in order. We can moreover generalize the integral equation 
(\ref{eq:fujii}) as follows. 
\begin{equation}
\label{eq:more-fujii}
\int\int{\cdots}\int_{{\bf R}^{n}}u(x-{\bf y}^{t}A{\bf y})dy_{1}dy_{2}{\cdots}dy_{n}
=f(x)
\end{equation}
where ${\bf y}=(y_{1},y_{2},\cdots,y_{n})^{t}$ and $A$ is positive--definite. 
However, calculation will be left to readers.

\vspace{5mm}
In the note we gave a generalization of the Lamb--Bateman integral equation 
and solved it in terms of fractional derivatives.  We can say that theory of 
fractional derivatives (more generally, {\bf Fractional Calculus}) is a powerful 
tool for solving integral equations.


\end{document}